\documentclass{book}
\usepackage[
 	lang = british, 
 	mode = work, 
]{emsproc}

\emsauthor{1}{James H. Davenport}

\authormark{James H. Davenport}

\emsaffil{1}{Department of Computer Science, University of Bath, Bath U.K.\\
\email{J.H.Davenport@bath.ac.uk}}

\def\R{{\mathbf R}}
\def\Z{{\mathbf Z}}

\theoremstyle{plain}

\theoremstyle{definition}

\begin{document}
\mainmatter


\chapter[Examples in Maths]{Digital Collections of Examples in Mathematical Sciences}


\keywords{Benchmarking, Citation, OpenMath}
\subjclass{{\bf 00A35}, 12-04, 20-04}



\begin{abstract}
Some aspects of Computer Algebra (notably Computation Group Theory and Computational Number Theory) have some good databases of examples, typically of the form ``all the $X$ up to size $n$''. But most of the others, especially on the polynomial side, are lacking such, despite the utility they have demonstrated in the related fields of SAT and SMT solving. We claim that the field would be enhanced by such community-maintained databases, rather than each author hand-selecting a few, which are often too large or error-prone to print, and therefore difficult for subsequent authors to reproduce.
\end{abstract}
\section{Introduction}
Mathematicians have long had useful collections, either of systematic data or examples. One of the oldest known such is the cuneiform tablet known as Plimpton 322, which dates back to roughly 1800BC: see \cite[pp. 172-176]{ConwayGuy1996}, or a more detailed treatment in \cite{Mansfield2021a,Robson2001}. This use of systematic tables of data spawned the development on logarithmic, trigonometric and nautical tables: Babbage's Difference Engine was intended to mechanise the production of such tables.
But there were also tables of purely mathematical interest: the author recalls using an 1839 table of  logarithms and  what are now known as Zech logarithms \cite{Zech1849} (but in fact they go back at least to \cite{Leonelli1803}), i.e. tables of the function $\log x \mapsto \log(1+x)$, at least over $\R$: Jacobi's table \cite{Jacobi1839} was modulo $p^n$ for all the prime powers $p^n<1000$.
\subsection{Data Citation}
Citation and referencing is an important point of modern scholarship --- Harvard-style referencing is generally attributed to \cite{Mark1881}, and the history of \emph{Science Citation Index} is described in \cite{Garfield2007}. It is well-understood, and practically all research students, and many undergraduates, get lessons in article citation practices.
\par
\begin{figure}[h]
	\caption{Overlaps between data citation harvesters \cite[Figure 5]{vandeSandtetal2019a}\label{Fig:F5}}
\includegraphics[scale=0.75]{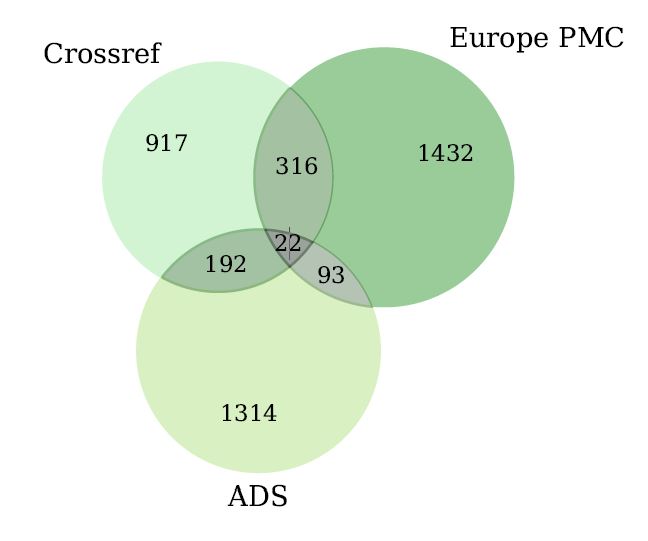}
\end{figure}
Despite the success of article citation, data citation is
a mess in practice \cite{vandeSandtetal2019a}: only 1.16\% of dataset DOIs in Zenodo are cited\footnote{In contrast, 60\% of papers in Natural Science and Engineering \emph{had} a citation in the next two years \cite{Lariviereetal2009,Remler2014a}.} (and 98.5\% of these are self-citations). It is  still a subject of some uncertainty: \cite{MooneyNewton2012a,KratzStrasser2014a} and significant
changes are still being proposed \cite{Daquinoetal2020a}.
Worse, perhaps, it is poorly harvested: see Figure \ref{Fig:F5}.
Assuming independence and looking at the overlap statistics, we can estimate that there are between 4,000--20,000 data sets waiting to be cited.
In such circumstances, {\it de facto\/} people cite a paper if they can find one.
\section{Pure Mathematics}
\subsection{Online Encyclopedia of Integer Sequences}
This database \cite{Sloane2003} can be said to have ``colonised the high ground'' in mathematics: mathematicians from all sub-disciplines use it. It has evolved from a private enterprise, for a long time at \verb!http://www.research.att.com/~njas/sequences!, to a system maintained by a foundation, and now at \url{https://oeis.org/}. The recommended citation is ``OEIS Foundation Inc. (2022), The On-Line Encyclopedia of Integer Sequences, published electronically at \url{https://oeis.org}, [date]'', but the author had originally to search the website to find it!
\subsection{Group Theory}
The Classification of Finite Simple Groups, as well as being a \emph{tour de force} in mathematics, also means that we have a complete database here. In most other areas, we have to be content with ``small' databases.
\par
An example of this is the transitive groups acting on $n$ points, where various authors have contributed: \cite{ButlerMcKay1983} ($n\le11$); \cite{Royle1987} ($n=12$); \cite{Butler1993} ($n=14,15$); \cite{Hulpke1996} ($n=16$); \cite{Hulpke2005} ($17\le n\le 31$); \cite{CannonHolt2008} ($n=32$).
These are available in the computer algebra system GAP (and MAGMA), except that (for reasons of space) $n=32$ isn't in the default build for GAP.
\par These are really great resources (if that's what you want), but how does one cite this resource: ``\cite[{\tt transgrp} library]
{GAP2021a}''?
\par
There are several other libraries such as primitive groups. But it could be argued that  (finite)
Group Theory is ``easy'': for a given $n$ there are a finite number and we ``just'' have to list them.
\subsection{$L$-functions and Modular Forms}
The $L$-functions and Modular Forms Database, known as LMFDB and hosted at \url{lmfdb.org} is a third example of mathematical databases.  The recommended citation, ``The LMFDB Collaboration, The L-functions and modular forms database, \url{http://www.lmfdb.org}, 2021'' is directly linked from the home page, which is a good model to follow.
\par
Computation in this area had a long history, from \cite{BirchSwinnertonDyer1963} and \cite{SwinnertonDyeretal1975} to the current database, which is the work of a significant number of people.  The early computations gave rise to the Birch--Swinnerton-Dyer Conjectures \cite{BirchSwinnertonDyer1965}, now a Clay Millennium Prize topic. The current computations are in active use by mathematicians: see Poonen's remarks in \cite{Davenportetal2018e}.
\section{SAT and SMT Solving}
\subsection{SAT Solving}
SAT solving is  normally seen as solving a Boolean expression written in Conjunctive Normal Form (CNF).

The 3-SAT problem is: given a 3-literals/clause CNF satisfiability problem, 
\begin{equation}\label{eq:1}
\underbrace{(l_{1,1}\lor l_{1,2}\lor l_{1,3})}_{\hbox{Clause 1}}\land (l_{2,1}\lor l_{2,2}\lor l_{2,3})\land\cdots\land (l_{N,1}\lor l_{N,2}\lor l_{N,3}),
\end{equation}
where $l_{i,j}\in\{x_1,\overline{x_1},x_2,\overline{x_2},\ldots\}$, is it satisfiable?  In other words, is there an assignment of $\{T,F\}$ to the $x_i$ such that all the clauses are \emph{simultaneously} true.
\par
3-SAT is the quintessential NP-complete problem \cite{Cook1966}.  2-SAT is polynomial, and $k$-SAT for $k>3$ is polynomial-transformable into 3-SAT.  In practice we deal with SAT --- i.e. no limitations on the length of the clauses and no requirement that all clauses have the same length.
\par
Let $n$ be the number of $i$ such that $x_i$ (and/or $\overline{x_i}$) actually occur.  Typically $n$ is of a similar size to $N$.
\par
Despite the problem class being NP-complete, nearly all examples are easy (e.g. SAT-solving has been routinely used in the German car industry for over twenty years \cite{KuchlinSinz2000}): either easily solved (SAT) or easily proved insoluble (UNSAT). For random problems there seems to be a distinct phase transition between the two: \cite{GentWalsh1994,AchlioptasPeres2004,AchlioptasPeres2006}, with the hard problems typically lying on the boundary.
\par
This means that constructing difficult examples is itself difficult, and a topical research area: \cite{Spence2015a,BalyoChrpa2018a}.
\par
SAT solving has many applications, so we want effective solvers for ``real'' problems, not just ``random'' ones.
This gives us the fundamental question: what does this mean?
\subsection{SAT Contests} 
These are described at \url{http://www.satcompetition.org}. They have 
been run since 2002. In the early years, there were distinct tracks for Industrial/Handmade/Random problems: this has been abandoned.
\par
The methodology is that the organisers accept submissions (from contestants\footnote{In 2020, contestants were required to submit at least 20 problems, as well as a solver.} and others), then produce a list of problems (in DIMACS, a standard format) and set a time (and memory) limit, and see how many of the problems the submitted systems can solve on the contest hardware.
\par
SAT is easy to certify (the solver just produces a list of values of the $x_i$). Verifying  UNSAT is much harder, but since 2013 the contest has required proofs of UNSAT for the UNSAT track,  and since 2020 in all tracks, in DRAT: a specified format (some of these proofs have been $>100$GB).
\par
The general feeling is that these contests have really pushed the development of SAT solvers, roughly speaking $\times2$/year. 
 For comparison, Linear Programming has done $\times1.8$ over a greater timeline and with more rigorous dcoumentation \cite{Bixby2015a}.
\subsection{SMT: Life Beyond SAT}
\begin{figure}[h]
	\caption{Available logics (March 2022) \url{https://smtlib.cs.uiowa.edu/logics.shtml}\label{Fig:SMT}}
\includegraphics[scale=0.46]{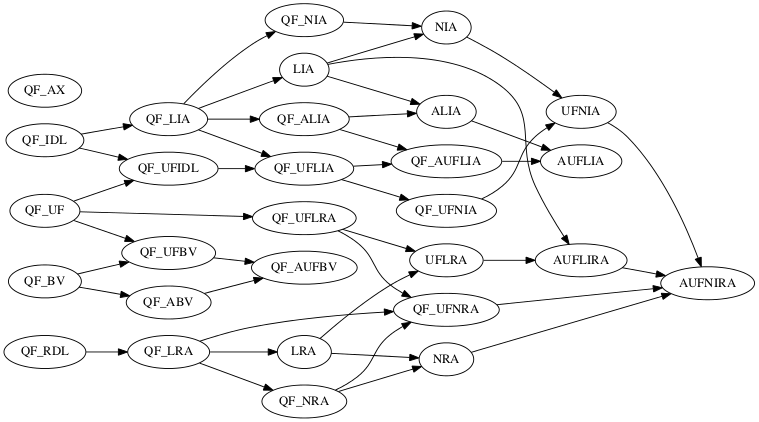}
\end{figure}

Consider a theory $T$, with variables $y_j$, and various Boolean-valued statements in $T$ of the form $F_i(y_1,\ldots,y_n)$, and a  CNF $\cal L$ in the form of (\ref{eq:1}) with $F_i(y_1,\ldots,y_n)$ rather than just $x_i$. In principle $T$ can be anything: those currently supported\footnote{By the SMT-LIB standard: \cite{Barrettetal2021a}, which also says `` New logics are added to the standard opportunistically, once enough benchmarks are available''.} are given in Figure \ref{Fig:SMT}.
\par
For example \verb+QF_NRA+ is the Quantifier-Free theory of Nonlinear Real Arithmetic, and \verb+QF_LRA+ (Linear Real Arithmetic) is included in this. Both \verb+QF_NRA+ and \verb+QF_UFLRA+ (Uninterpreted Functions and Linear Real Arithmetic) are included in \verb+QF_UFNRA+.
\par
Then the SAT/UNSAT question is similar: do there exist values of $y_i$ such that  $\cal L$ is true (SAT), or can we state that no such exist (UNSAT), and the community runs SMT Competitions (\url{https://smt-comp.github.io/2022/}). There is a separate track for each theory $T$, as the problems will be different. Within each, the problems are subdivided as industrial/crafted/random.
\par
The SMT-LIB format  \cite{Barrettetal2021a} provides a standard input format. The question of proving UNSAT is in general unsolved (but see \cite{Kremeretal2021a} for one particular theory $T$).
\par
There has been substantial progress in SMT-solving over the years, possibly similar to SAT, and probably also spurred by the contests.
\section{Computer Algebra: Where are we?}
Obviously, Group Theory and others are parts of computer algebra: what about the rest of computer algebra?
\par
\emph{In general} the problems of computer algebra have a bad worst-case complexity, and we want effective solvers for ``real'' problems, not just ``random'' ones.  The question, as in SAT and SMT, is ``what does this mean?''.

But there are also various logistical challenges.
\begin{enumerate}
\item{\bf Format:} there is no widely accepted common standard. We do have OpenMath \cite{Buswelletal2017a}, but it's not as widely supported as we would like.
\item{\bf Contests:} There are currently none.  Could SIGSAM organise them?
\item{\bf {Pr}oblem Sets:} There are essentially no independent ones. Each author chooses his own.
\item{\bf Archive:} Not really.
\end{enumerate}
We now consider various specific problems.
\subsection{Polynomial GCD}
This problem is NP-hard (for sparse polynomials, even univariate)
\cite{Plaisted1984,DavenportCarette2010}. Even for dense polynomials, it 
 can be challenging for multivariates.
There is no standard database: one has to trawl previous papers  (and often need to ask the authors, as the polynomials were too big to print in the paper).
Verification is a challenge: one can check that the result is \emph{a common divisor}, but verifying \emph{greatest} is still NP-hard \cite{Plaisted1984}.
\subsection{Polynomial Factorisation}
This is known to be polynomial-time for dense encodings \cite{Lenstraetal1982}, even though their exponent is large, and much work has gone into better algorithms, e.g. \cite{Abbottetal2000a}. Presumably it is NP-hard for sparse encodings, though the author does not know of an explicit proof.
There is no standard database: one has to trawl previous papers  (and often need to ask the authors, as the polynomials were too big to print in the paper).
\par Verification is a challenge: one can check that the result is \emph{a factorisation}, but checking completeness (i.e. that these factors are irreducible) seems to be as hard as the original problem in the worst cases.
\par
It is worth noting that, with probability 1, a random dense polynomial is irreducible (and easily proved so by the Musser test \cite{Musser1978}), so the question ``what are the \emph{interesting} problems?'' is vital.
\subsection{Gr\"obner Bases}
The computation of Gr\"obner bases has many applications, from engineering to cryptography. But this has doubly exponential (w.r.t. $n$, the number of variables) worst-case complexity \cite{MayrRitscher2013a}, even for a prime ideal \cite{Chistov2009}.
If we take $n$ ``random'' equations in $n$ variables, they will satisfy the conditions for the Shape Lemma \cite{Beckeretal1994} and have $D\le n^n$ solutions, so a Gr\"obner base in a purely lexicographical order will look like 
\begin{equation}\label{eq:GB}
\{p_1(x_1), x_2-p_2(x_1),x_3-p_3(x_1),\ldots,x_n-p_n(x_1)\},
\end{equation}
where $p_1$ is a polynomial of degree $D$ in $x_1$ and the other $p_i$ are polynomials of degree at most $D-1$ in $x_1$. Experience shows that the coefficients of the $p_i$ will generally be large (theoretically, they can be $D$ times as long as the input coefficients).  Conversely, if we have $n+1$ equations, there are generally no solutions and the Gr\"obner base is $\{1\}$: much shorter than (\ref{eq:GB}).

The good news from the point of view of this paper is that there is a collection \cite{BiniMourrain1996}, but it's very old (1996), so most of the examples are trivial with today's  hardware and software, and completely static. Worse, some of the examples are only available in PDF. 

There always is a Gr\"obner base (no concept of UNSAT as such) but it's not clear what a useful certificate of ``$G$ is a Gr\"obner base for input $L$'' might mean in general (but see \cite{Arnold2003}). If $G=\{g_1,\ldots,g_M\}$ is a Gr\"obner base of $F=\{f_1,\ldots,f_N\}$ then a general certificate would consist of three components:
\begin{enumerate}
\item A proof that $G$ is a Gr\"obner base, which would mean that every $S$-polynomial $S(g_i,g_j)$ reduces to 0 under $G$, which is easily checked;
\item A proof that $(F)\subseteq(G)$, which could be a set of $\lambda_{i,j}$ such that every $f_i=\sum\lambda_{i,j}g_j$;
\item A proof that $(G)\subseteq(F)$, which could be a set of $\mu_{i,j}$ such that every $g_i=\sum\mu_{i,j}f_j$.
\end{enumerate}
However, the $\lambda_{i,j}$ and $\mu_{i,j}$ might be (and generally are) extremely large.
\subsection{Real Algebraic Geometry}

Again, the problem of describing the decomposition of $\R^n$ sign-invariant for a set $S$ of polynomials $f_i$ in $n$ variables has doubly exponential (w.r.t. $n$) worst-case complexity \cite{BrownDavenport2007}. However, unlike Gr\"obner bases, it seems that this is the ``typical'' complexity, though the author knows no formal statement of this. For a given problem, the complexity can vary greatly: \cite[Theorem 7]{BrownDavenport2007} is an example of a polynomial $p$ in $3n+4$ variables such that \emph{any} Cylindrical Algebraic Decomposition (CAD), w.r.t. one order, of $\R^{3n+4}$ sign-invariant for $p$ has $O\left(2^{2^n}\right)$ cells, but w.r.t. another order has 3 cells.
\begin{eqnarray*}
&p:=x^{n+1}\left(\left(y_{n-1}-\frac12\right)^2+\left(x_{n-1}-z_n\right)^2\right)\left(\left(y_{n-1}-z_n\right)^2+\left(x_{n-1}-x_n\right)^2\right)\\
&+\sum_{i=1}^{n-1}x^{i+1}\left(\left(y_{i-1}-y_i\right)^2+\left(x_{i-1}-z_i\right)^2\right)\left(\left(y_{i-1}-z_i\right)^2+\left(x_{i-1}-x_i\right)^2\right)\\
&+x\left(\left(y_{0}-2x_0\right)^2+\left(\alpha^2+(x_0-\frac12)\right)^2\right)\times\\&\left(\left(y_{0}-2+2x_0\right)^2+\left(\alpha^2+(x_0-\frac12)\right)^2\right)+a.
\end{eqnarray*}
The bad order (eliminating $x$, then $y_0,\alpha,x_0,z_1,y_1,z_1,\ldots$, $x_n,a$) needs $O\left(2^{2^n}\right)$ (Maple: 141 when $n=0$) cells.
Any order eliminating $a$ first says that $R^{3n+3}$ is undecomposed, and the only question is $p=0$, which is linear in $a$, and we get three cells: $p<0$, $p=0$ and $p>0$.

However, if we replace $a$ by $a^3$, the topology is essentially the same, but the discriminant is no longer trivial, and the ``good'' order now generateses 213 cells in Maple, rather than three.

There is a collection \cite{Wilsonetal2012b}, not quite as old as \cite{BiniMourrain1996} (2014 was the last update), but still completely static.
The DEWCAD project \cite{Bradfordetal2021a} might update this, but there are still issues of long-term conservation. The format has learned from \cite{BiniMourrain1996} and each example is available in text, Maple input and QEPCAD.

If we are just looking at computing a CAD, which we might wish to do for motion planning purposes \cite{Wilsonetal2013c}, there is no concept of UNSAT, and the question of certificates of correctness is essentially unsolved. Attempts to produce a formally verified CAD algorithm have also so far been unsuccessful \cite{CohenMahboubi2010}.

However, CAD was invented \cite{Collins1975} for the purpose of quantifier elimination, i.e. converting $Q_kx_kQ_{k+1}x_{k+1}\cdots Q_nx_n\Phi(f_i)$, where $Q_i\in\{\exists,\forall\}$ and $\Phi$ is a Boolean combination of equalities and inequalities in the $f_i$, into $\Psi(g_1,\ldots,g_{n'})$, where $\Psi$ is a Boolean combination of equalities and inequalities in the $g_i$, polynomials in $x_1,\ldots,x_{k-1}$, and if the statement is fully quantified, the result is a Boolean. A common case, particularly in program verification, is the fully existential case (all $Q_i$ are $\exists$), where $\Phi$ is  ``something has gone wrong'', and we want to show this can't happen. Then SAT is easy (exhibit values of $x_i$ such that $\Phi$ is true,  but UNSAT is much harder to certify. See \cite{Kremeretal2021a} for some steps in this direction.
\subsection{Integration}
The computational  complexity of integration, i.e. given a formula $f$ in a class $\cal L$, is there a formula $g\in\cal L$, or in an agreed extension of $\cal L$, such that $g'=f$, is essentially unknown (but integration certainly involves GCD, factorisation etc.). When $\cal L$ includes algebraic functions, difficult questions of algebraic geometry arise (see \cite[as corrected in \cite{MasserZannier2020b}]{Davenport1981a}), and there is no known bound on the complexity of these.

 ``Paper'' mathematics produced large databases of integrals,
 e.g. \cite{GradshteynRyzhik2007}, but these are (at best) in PDF, and the way they are commonly printed makes it extremely hard to recover semantics from the layout.
Probably the best current database is described in \cite{JeffreyRich2010}. But these databases are almost entirely of successful (SAT in our notation) examples, and there is almost no collection of UNSAT ($\not\!\exists g\in{\cal L}: g'=f$) examples.
Algorithm-based software (e.g. \cite{Davenport1981a}) has an internal proof of UNSAT, but I know of no software that can exhibit it. That proof is typically very reliant on the underlying mathematics.

 A new question here is the ``niceness'' of the output in the SAT case. Jeffrey and Rich \cite{JeffreyRich2010} give the example of
\begin{equation}\label{eq:I1}
\int{\frac{5x^4}{(1+x)^6}{\rm d}x}=\frac{x^5}{(1+x)^5},
\end{equation}
where Maple's answer is
\begin{equation}\label{eq:I2}
\frac{-10}{\left( 1+x \right) ^{3}}+\frac{5}{\left( 1+x \right) ^{4}}-\frac{5}
	{\left( 1+x \right)}- \frac1{\left( 1+x \right) ^{5}}+\frac{10}{\left( 1+x
	\right) ^{2}}.
\end{equation}
Note that (\ref{eq:I2}) is not just an ugly form of the right-hand side of (\ref{eq:I1}): the two differ by 1, which is a legitimate constant of integration.
\par
While some element of ``niceness'' is probably beyond automation, ``simplicity'' in the sense of \cite{Carette2004}, essentially minimal Kolmogorov complexity, is probably a good proxy, and could be automatically judged (at least in principle: there are probably some messy system-dependent issues in practice).
\section{Conclusions}
\begin{enumerate}
\item The field of computer algebra really ought to invest in the sort of contests that have stimulated the SAT and SMT worlds.
\item This requires much larger databases of ``relevant'' problems than we currently have, and they need to be properly curated.
\item[+]The technology of collaborative working, e.g. wikis, or GitHub, has greatly advanced since the days of \cite{BiniMourrain1996}, which should make collaborative construction of example sets easier, and would also help with the preservation challenge.
\item[--]Although OpenMath is in principle a suitable system-neutral notation that could be the standard input (and output) format, such a use would challenge OpenMath implementations. This would be a good development, though.
\item This would allow much better benchmarking practices: see the description in \cite{Brainetal2017a}.
\item There are significant challenges in providing ``certificates'', not just of UNSAT in the case of integration, but elsewhere in algebra. For example, asserting $g=\gcd(f_1,f_2)$ involves, not just the claim that $g$ divides $f_1$ and $f_2$, but also that $f_1/g,f_2/g$ are relatively prime, which may be much harder to demonstrate.
\end{enumerate}


\begin{ack}
The author is grateful to Dr.~Uncu for his comments on drafts, and to the organisers of the MIDAS session at the 8th European Congress of Mathematicians for prompting these reflections.
\end{ack}

\begin{funding}
This work was partially supported by EPSRC Grant  EP/T015713/1.
\end{funding}


\bibliographystyle{emsplain}

\end{document}